\documentclass[lettersize,journal]{IEEEtran}
\usepackage{amsmath,amsfonts}
\usepackage{array}
\usepackage{textcomp}
\usepackage{stfloats}
\usepackage{url}
\usepackage{verbatim}
\usepackage{graphicx}
\usepackage{algorithm}
\usepackage{caption}
\usepackage{subcaption}
\usepackage[noend]{algpseudocode}
\usepackage[table]{xcolor}

\def\BibTeX{{\rm B\kern-.05em{\sc i\kern-.025em b}\kern-.08em
    T\kern-.1667em\lower.7ex\hbox{E}\kern-.125emX}}
\usepackage{balance}
\begin{document}
\title{Automatic design optimization of preference-based subjective evaluation with online learning in crowdsourcing environment}
\author{Yusuke Yasuda and Tomoki Toda}

\markboth{Journal of \LaTeX\ Class Files,~Vol.~18, No.~9, September~2020}%
{How to Use the IEEEtran \LaTeX \ Templates}

\maketitle
\begin{abstract}
A preference-based subjective evaluation is a key method for evaluating generative media reliably. However, its huge combinations of pairs prohibit it from being applied to large-scale evaluation using crowdsourcing. To address this issue, we propose an automatic optimization method for preference-based subjective evaluation in terms of pair combination selections and allocation of evaluation volumes with online learning in a crowdsourcing environment. We use a preference-based online learning method based on a sorting algorithm to identify the total order of evaluation targets with minimum sample volumes. Our online learning algorithm supports parallel and asynchronous execution under fixed-budget conditions required for crowdsourcing. Our experiment on preference-based subjective evaluation of synthetic speech shows that our method successfully optimizes the test by reducing pair combinations from 351 to 83 and allocating optimal evaluation volumes for each pair ranging from 30 to 663 without compromising evaluation accuracies and wasting budget allocations.
\end{abstract}
\section{Introduction}
\label{sec:intro}
Subjective evaluations are essential for evaluating generative media, such as synthetic speech. Recent advancements in generative models are expected to make subjective evaluation difficult \cite{Shirali2018}. For example, some text-to-speech synthesis methods reached naturalness close to human speech in simple read-a-loud tasks \cite{Shen2017, DBLP:journals/corr/abs-1809-08895, DBLP:conf/icml/KimKS21}. Thus, modern subjective evaluation methods are required to distinguish small quality differences. The mean opinion score (MOS) is the standard subjective evaluation method for synthetic speech that rates samples by direct scores based on quality scales. MOS is widely used, but its issues on various biases are also thoroughly studied \cite{zielinski2008on, zielinski2016on}. Under the current situation of generative technologies, the limitations of MOS evaluation are being clarified. The limitations of MOS result in the following treatments of MOS. First, it behaves as a relative quality score, so regarding it as absolute quality is dangerous \cite{LEMAGUER2024101577}. Second, MOS from different experiments must not be compared \cite{cooper2023investigating}. Third, many evaluations and evaluators should be collected to offset the bias of MOS \cite{rosenberg17_interspeech}. In reality, the cost of crowdsourcing limits the collectible number of evaluations. As a result, it is suspected that MOS is saturated, and quality differences in MOS have reached the limit where the confidence interval can not be improved by increasing evaluation volume considering the cost of crowdsourcing \cite{yasuda23_interspeech}.

An alternative subjective evaluation method to the direct-scoring-based evaluation such as MOS is preference-based evaluation, which rates two samples by selecting the better one. The preference-based test has good properties compared to MOS. 
The comparative scoring is expected to rate quality difference more clearly than direct scoring because it has clear references for evaluators to rate without depending on ambiguous score scales \cite{DBLP:journals/corr/ShahBBPRW14}. Thus, it is expected to provide more reliable scores with less evaluation volume. 
In addition, preference-based evaluation is expected to provide consistent scores across different experiments.
On the other hand, the essential issue of preference-based subjective evaluation is the cost caused by a huge combination of pairs. Due to the huge combination of pairs, preference-based subjective evaluation can not be applied to large-scale evaluations containing many evaluation targets. This is why a preference-based test is usually conducted after MOS evaluation by selecting a few pairs with similar scores. However, considering the cost of crowdsourcing, the two-phased subjective evaluations are redundant and inefficient. 

In this research, we automatically optimize the preference-based subjective evaluation by dynamically designing pair combinations to be evaluated and allocation of evaluation volumes for each pair. We use online learning to optimize preference-based tests \cite{DBLP:journals/jmlr/BengsBMH21}. Online learning can optimize evaluation accuracy or budget by optimal allocation of evaluation volumes based on estimating scores of evaluation targets online. This feature of online learning enables us to apply preference-based tests to large-scale evaluations with realistic costs. On the other hand, it is not straightforward to execute online learning in a crowdsourcing environment because online learning is a sequential algorithm with unknown exact convergence complexity.
In contrast, crowdsourcing requires parallel and asynchronous execution under a fixed budget. To address the difficulties of online learning in a crowdsourcing environment, we propose a stable online learning algorithm that supports parallel and asynchronous execution under fixed budget settings. We show the effectiveness of our algorithm in an experiment on the subjective evaluation of synthetic speech in a crowdsourcing environment.

Our contributions are summarized as follows:
\begin{itemize}
    \item We optimize pair selection of preference-based subjective evaluations with online learning to achieve high evaluation accuracy under a limited budget.
    \item We propose methods stabilizing online learning in asynchronous and parallel execution to apply it to a large-scale evaluation using crowdsourcing.
    \item Our experiment on preference-based subjective evaluation reduces 351 total combinations to 83 pairs to determine the total order of qualities of speech synthesis systems.
    \item Our method achieved more pairs with statistically significant quality differences than MOS by avoiding the contraction bias observed in the MOS evaluation.
\end{itemize}

\section{Issues in subjective evaluation}

MOS is the standard subjective evaluation method to evaluate synthetic speech. MOS is an evaluation method based on direct scoring with five-point scales ranging from 1: Bad to 5: Excellent. The MOS evaluation is widely used because it is easy to use and reliable compared to objective quality metrics. However, MOS is known to suffer from many biases. The biases observed in MOS include, for example, contraction, centering, range equalizing, and scale nonlinearity \cite{zielinski2008on}. The contraction bias is the compression of the score distribution. The centering bias is a systematic shift of all scores. The range equalizing bias is the spanning of scores to the entire scale. The scale nonlinearity bias is inconsistent differences in the semantics of the labels. Due to these biases, MOS should not be interpreted as an absolute quality score but a relative score \cite{LEMAGUER2024101577}. 

Another issue of MOS is the lack of rigorous statistical analysis methods. It is known that MOS does not follow normal approximation well. This is because the distribution of opinion scores is not Gaussian, and the typical sample volume to estimate MOS is too small to assume the central limit theorem. These situations suggest that the normal approximation error estimated by Berry–Esseen theorem \cite{10.2307/1990053} is large when we analyze MOS assuming the central limit theorem. Therefore, a nonparametric statistical test such as a rank test is recommended to check the hypothesis whether one is better than the other \cite{rosenberg17_interspeech}. Nevertheless, the confidence interval for MOS is usually reported based on the central limit theorem. Naturally, underestimating the confidence interval in MOS is concerned \cite{yasuda23_interspeech, camp23_interspeech}. 

MUSHRA is a subjective evaluation scheme that aims to reduce biases observed in MOS. MUSHRA includes a hidden reference and a low-quality anchor in an evaluation set. The MUSHRA test can avoid the contraction and scale nonlinearity biases, but it still suffers from the centering and range equalizing biases. In addition, the lack of rigorous statistical analysis methods is also an issue for MUSHRA.

A preference-based test is a simple evaluation method that selects one out of two options better. There are two advantages of the preference-based test \cite{zielinski2008on}. The first advantage is that it can avoid various biases. Preference-based tests use scale-free binary scores for two samples that work as references to each other. This feature of the preference-based test enables us to avoid the contraction, centering, range equalizing, and scale nonlinearity biases, which are typically observed in MOS. The second advantage is that rigorous statistical methods can be used to analyze the results of preference-based tests. The preference scores can be considered samples from the Bernoulli distribution, and the win counts of Bernoulli samples follow the binomial distribution. This knowledge of probability distribution enables us to use statistical analysis without relying on normal approximation. For example, the binomial test can be used as a statistical test for preference-based tests to check the hypothesis whether one is better than the other. In addition, the Clopper–Pearson confidence interval can be used for preference-based tests to calculate confidence intervals that reflect the asymmetric nature of binomial distribution without relying on the central limit theorem \cite{be7c0fd0-f562-39ad-b8e0-716a276561d1}. These methods work reliably for a small number of samples from a binomial distribution where errors of normal approximation by the central limit theorem are expected to be large.

The main disadvantage of preference-based tests is their large evaluation cost due to the huge combination of pairs. If there are $|S|$ evaluation targets, the order of the pair combination to evaluate is $O(|S|^2)$. This huge combination of pairs prevents us from applying a preference-based test to large-scale evaluations, although it has huge benefits for scientific research that requires reliability and accuracy of evaluations.

\section{Dynamic optimization of preference test}
\subsection{An idea of dynamic preference test}
\label{subsec:dpt}

\begin{figure}[t]
    \begin{center}
    {\includegraphics[width=1.0\columnwidth]{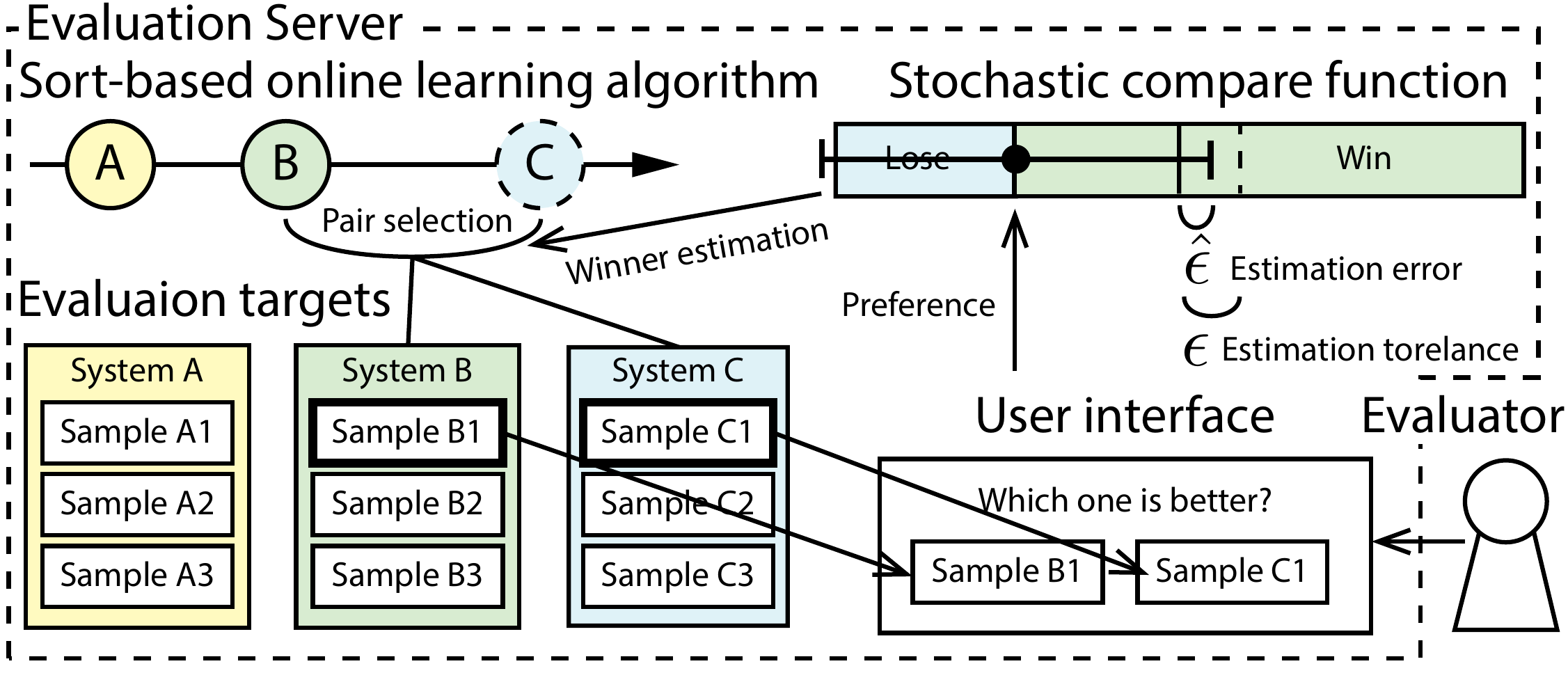}}
    \end{center}
    \caption{An overview of dynamic preference test.}
    \label{fig:overview}
\end{figure}

Figure \ref{fig:overview} shows an overview of our approach for the dynamic preference test. A preference test is a subjective evaluation method based on selecting a better one out of two. The issue of the preference test is a huge combination of pairs: the order of pairs is $O(|S|^2)$ where $|S|$ is the number of evaluation targets $S$. This contrasts value-based evaluation methods such as MOS, which require evaluating $O(|S|)$ targets. On the other hand, there is room for intuitive optimization of the preference test. For example, we usually do not test all combinations of pairs by preference tests. We usually use preference tests to evaluate only part of targets with similar quality. This convention is because we assume that target A would be preferred to C based on the positive preference between A and B and the positive preference between B and C. A preference test for A and C can be skipped in this case. In addition, we usually reduce the evaluation budget for targets with distinct quality, and we increase the evaluation budget for targets with similar quality. The quality estimation of the pair selection and budget allocation is usually based on a preliminary informal listening test to design a preference test. These conventions indicate that we could optimize pair combination and budget allocation of preference tests if we could estimate the quality of targets during an evaluation.

Online learning is a score estimation method for evaluation targets with unknown scores that can optimize evaluation volumes for the score estimation as little as possible \cite{DBLP:journals/jmlr/BengsBMH21}. There are two approaches of online learning distinguished with different objectives: total score maximization under a limited budget or score identification with the desired accuracy. For an application to subjective evaluation, we need online learning for the identification approach because we aim to evaluate synthetic media quality with high accuracy and efficiency. We, therefore, adopt an online learning method for preference identification to optimize the design of preference-based subjective evaluation. For this purpose, sort-based online learning is suitable because it can optimize pair combinations by finding pairs with similar scores with minimum evaluations to identify the total quality order of evaluation targets.

\subsection{MERGE-RANK algorithm}
MERGE-RANK algorithm (MRA) is a sort-based online learning method for identifying the total order of evaluation targets based on preference-based subjective evaluation \cite{DBLP:conf/icml/FalahatgarOPS17}. MRA optimizes the combination of pairs by comparing the minimum pairs required to determine the total order of evaluation targets. To do so, MRA performs a merge sort algorithm on evaluation targets by estimating their true preference based on the preference scores submitted by evaluators. MRA also optimizes the number of evaluations required to estimate the true preference for each pair by using evaluation termination criteria based on a statistical test. The two optimization features of MRA, i.e. the optimal pair combinations and evaluation volume allocation, can realize the automatic pair selection with similar scores and minimization of total evaluation volumes to achieve the desired evaluation accuracy for a preference test.

Algorithm \ref{alg:mergerank} shows MRA. MRA is based on the merge sort algorithm \cite{goldstine1963planning}, which is a divide-and-conquer algorithm that recursively divides an unordered set into two ordered sets followed by merging them with MERGE algorithm as shown in Alg. \ref{alg:merge}. The only difference between the MRA and merge sort is that MERGE uses a stochastic COMPARE function based on the evaluator's preference scores instead of a deterministic one. The algorithm of COMPARE is shown in Alg. \ref{alg:compare}. COMPARE determines the winner from a comparing pair by checking if the mean of preference scores is smaller or greater than the even threshold. Let $i$ and $j$ be a comparing pair, $r_{ij}$ be the number of comparisons, $w_{ij}$ be the win counts of $i$ over $j$, and $\hat{p}_{ij}$ be the win rate of $i$. Then the win rate of $i$ is $\hat{p}_{ij} = \frac{w_{ij}}{r_{ij}}$. The winner can be determined by comparing $\hat{p}_{ij}$ with an even threshold of 0.5: $i$ is the winner if $\hat{p}_{ij} > \frac{1}{2}$. For evaluations of synthetic speech, the elements of comparing pair $i$ and $j$ represent speech synthesis systems. COMPARE randomly selects two synthetic speech samples generated from each speech synthesis system and presents them to evaluators. An evaluator submits a preference for the two speech synthesis systems to the COMPARE function by listening to the two speech samples.

\begin{algorithm}[t]
\caption{MERGE-RANK}\label{alg:mergerank}
\textbf{Input:} Set $S$, bias $\epsilon$, confidence $\delta$.
\begin{algorithmic}
\State $S_1 = \text{MERGE-RANK}(S(1:\lfloor|S/2|\rfloor),\epsilon,\delta)$
\State $S_2 = \text{MERGE-RANK}(S(\lfloor|S/2|\rfloor+1:|S|),\epsilon,\delta)$
\end{algorithmic}
\textbf{Output:} $\mathrm{MERGE}(S_1,S_2)$
\end{algorithm}
\begin{algorithm}

\caption{MERGE}\label{alg:merge}
\textbf{Input:} Sorted sets $S_1,S_2$, bias $\epsilon$, confidence $\delta$.\\
\textbf{Initialize:} $i = 1, j = 1$ and $O = \emptyset$.
\begin{algorithmic}
\While{$i \le |S_1|~\text{and}~j \le |S_2|$}
\If{$S_1(i) = \mathrm{COMPARE}(S_1(i),S_2(j),\epsilon,\delta)$}
\State append $S_2(j)$ at the end of $O$ and $j = j + 1$.
\Else
\State append $S_1(i)$ at the end of $O$ and $i = i + 1$.
\EndIf
\If{$i \le |S_1|$}
\State append $S_1(i:|S_1|)$ at the end of $O$.
\EndIf
\If{$j \le |S_2|$}
\State append $S_2(j:|S_2|)$ at the end of $O$.
\EndIf
\EndWhile
\end{algorithmic}
\textbf{Output:} Sorted set $O$
\end{algorithm}

\begin{figure}[!t]
    \begin{center}
    \begin{subfigure}[b]{1.0\columnwidth}
    {\includegraphics[width=1.0\columnwidth]{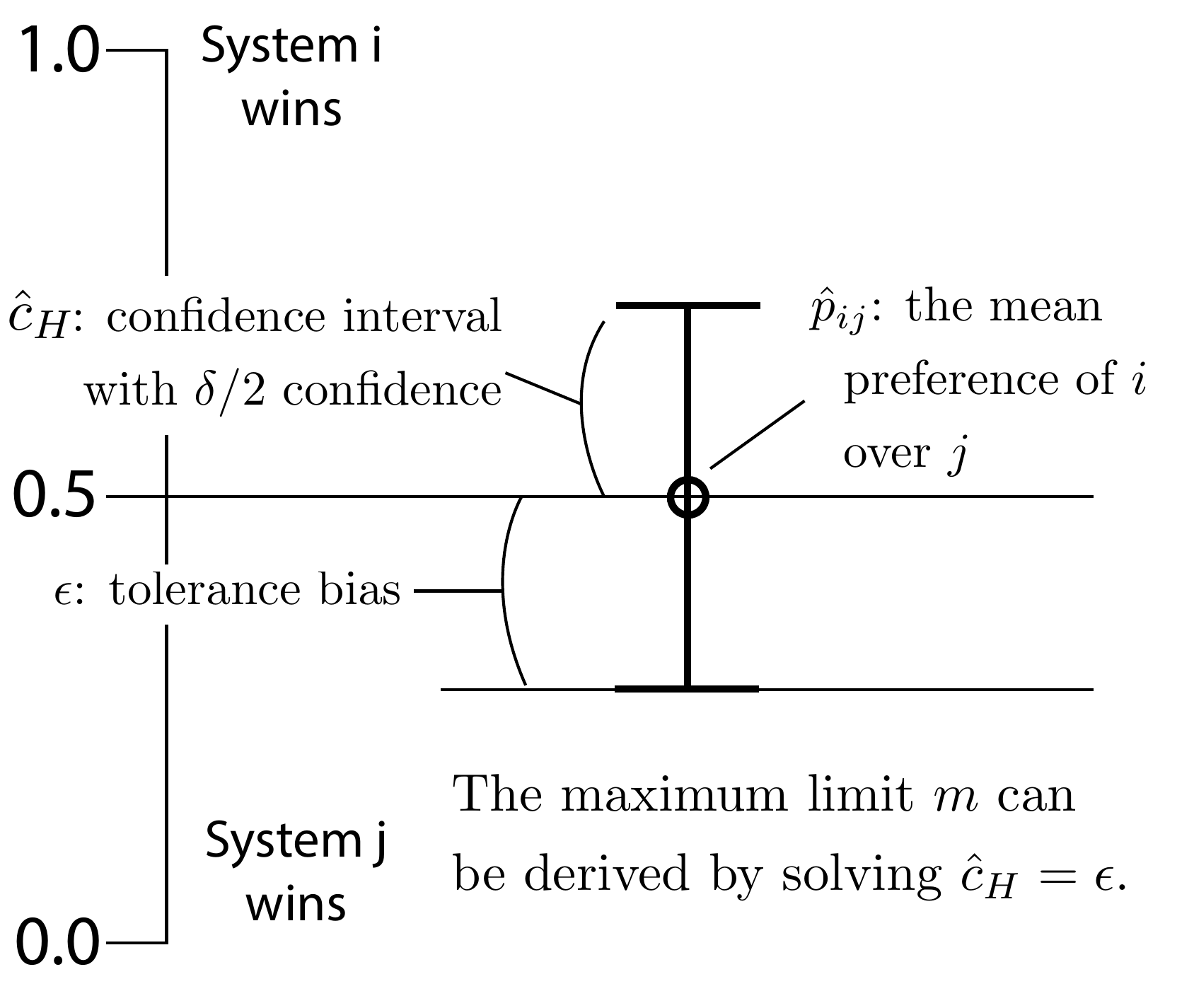}}
    \caption{The worst case where preference is a tie. This case requires the maximum number of evaluations $m$ to achieve tolerance bias $\epsilon$.}
    \label{fig:epsilon-worst}
    \end{subfigure}
    \end{center}
    \begin{center}
    \begin{subfigure}[b]{1.0\columnwidth}
    {\includegraphics[width=1.0\columnwidth]{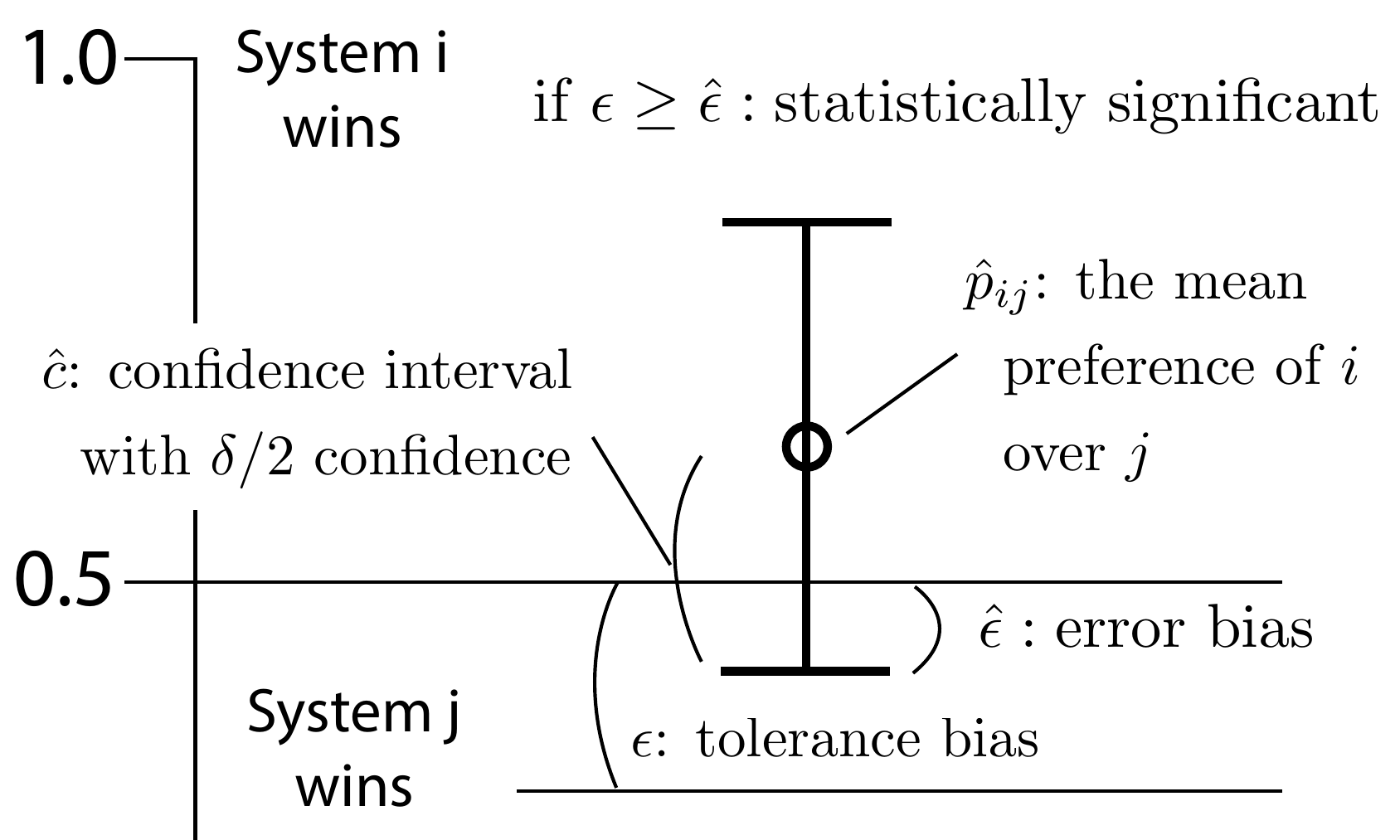}}
    \caption{An ideal case where an evaluation can be terminated early. In this case, the error bias $\hat{\epsilon}$ becomes less than tolerance bias $\epsilon$ before reaching maximum limit $m$.}
    \label{fig:epsilon-good}
    \end{subfigure}
    \end{center}
    \caption{The winner estimation based on the statistical test.}
    \label{fig:epsilon}
\end{figure}

The MRA ensures a specified evaluation accuracy for all pairs. COMPARE estimates the winner with at most confidence $\delta$ and tolerance bias $\epsilon$ as shown in Fig.~\ref{fig:epsilon}. To do so, COMPARE evaluates, at most, the maximum comparisons $m$. The worst case that requires maximum comparisons $m$ happens when preference is a tie, as shown in Fig.~\ref{fig:epsilon-worst}. In this case, confidence interval $\hat{c}_H$ must be less than or equal to tolerance bias $\epsilon$ to achieve the specified evaluation accuracy. Therefore, the maximum evaluation limit $m$ can be derived by solving $\hat{c}_H = \epsilon$. The confidence interval $\hat{c}_H$ can be derived from Hoeffding's inequality \cite{doi:10.1080/01621459.1963.10500830}, which can evaluate the upper bound of tail probability showing sample mean $\hat{p}$ from $r$ samples deviates from true mean $p$ by $\Delta$ as $\mathbb{P}[\hat{p} \le p - \epsilon] \le e^{-2r\Delta^2}$.
Based on the inequality, the confidence interval with confidence $\delta$ is then given by $\hat{c}_H \le \sqrt{1/(2r)\log(2/\delta)}$ from $\frac{\delta}{2} \le e^{-2r\hat{c}_H^2}$.
The maximum comparisons $m$ that ensures confidence $\delta$ with tolerance bias $\epsilon$ is obtained as  $m = \frac{1}{2\epsilon^2}\log\frac{2}{\delta}$ derived from
$\epsilon = \sqrt{1/(2m)\log(2/\delta)}$.

If a pair has a clear preference by a large margin, it is not necessary to evaluate up to the maximum comparisons to ensure confidence and tolerance bias. To optimize the number of comparisons further in such a case, COMPARE stops evaluation based on a statistical test as shown in Fig.~\ref{fig:epsilon-good}. Based on the confidence interval 
$\hat{c}(r)$
, COMPARE determines the winner if error bias $\hat{\epsilon}(r, \hat{p}) = \hat{c}(r) - |\hat{p} - \frac{1}{2}|$ becomes less than tolerance bias $\epsilon$. To justify the early termination where the winner may fluctuate unstably with small evaluation volume $r$, this stopping criterion must ensure that the estimated winner is stable during the whole evaluation history and that winner reversal only happens at most the error probability $\delta$ and error bias $\epsilon$. Therefore, the confidence interval $\hat{c}(r)$ used for the early termination is more conservative than the confidence interval $\hat{c}_H$ used for the maximum limit $m$. 
Let us assume COMPARE estimates $i$ as a winner at $r$ times comparisons. At each time, the tail probability misidentifying $i$ is not a winner is bounded based on Hoeffding's inequality by $\frac{\delta}{4r^2}$ as $\mathbb{P}(\hat{p}_{ij}^r - \frac{1}{2} > \hat{c}(r) - \epsilon) \le \exp(-2r(\hat{c}(r) - \epsilon)^2) \le \exp(-2r(\hat{c}(r))^2) = \exp(-\log\frac{4r^2}{\delta}) = \frac{\delta}{4r^2}$. After $1,\dots,r$ times comparison iteration, the tail probability of the misidentification is bounded by $\frac{\delta}{2}$ for any $r$ as $\mathbb{P}(\cup_{r=1}^\infty\{\hat{p}_{ij}^r - \frac{1}{2} > \hat{c}(r) - \epsilon\}) \le \sum_{r=1}^\infty \mathbb{P}(\hat{p}_{ij}^r - \frac{1}{2} > \hat{c}(r) - \epsilon) = \sum_{r=1}^\infty\frac{\delta}{4r^2} = \frac{\pi^2}{24}\delta \le \frac{\delta}{2}$, where $\hat{p}_{ij}^r$ is a win rate at $r$-time comparison. Here, we use subadditivity of probability $P(\cup_{n=1}^\infty A_n) \le \sum_{n=1}^\infty P(A_n)$ and $\sum_{r=1}^\infty\frac{1}{r^2} = \frac{\pi^2}{6}$. If we consider the opposite case where COMPARE estimates $j$ as a winner at $r$ comparisons, we obtain the error rate of COMPARE is at most $\delta$. 

MRA has bounded complexity to converge. MRA compares only a small subset of pairs from all combinations to converge, and the number of pairs to be compared is bounded. If the set $S$ contains $|S|$ evaluation targets, the number of compared pairs to converge $T(|S|)$ is bounded by the best and worst sorting complexity $\frac{|S|}{2}\log_2|S| \le T(|S|) \le |S|\log_2|S| - |S| + 1$ when $|S|$ is the power of two because it is based on merge sort. The order $O(|S|\log|S|)$ is much more efficient than that of the total combination $O(|S|^2)$. To sum up, MRA converges at most $m \times \max T(|S|)$ evaluations.

\begin{algorithm}[t]
\caption{COMPARE}\label{alg:compare}
\textbf{Input:} element pair $i,j$, bias $\epsilon$, confidence $\delta$.

\textbf{Initialize:} $\hat{p}_{ij} = \frac{1}{2},m=\frac{1}{2\epsilon^2}\log\frac{2}{\delta},r_{ij}=0,w_{ij}=0$. \\
\textbf{Define:} $\hat{c}(r) = \sqrt{\frac{1}{2r}\log\frac{4r^2}{\delta}}$ if $r > 0$ else $\frac{1}{2}$.\\
\textbf{Define:} $\hat{\epsilon}(r, \hat{p}) = \hat{c}(r) - |\hat{p} - \frac{1}{2}|$.
\begin{algorithmic}
\While{$\epsilon \le \hat{\epsilon}(r_{ij}, \hat{p}_{ij})$ and $r_{ij} \le m$}
\State Compare $i$ and $j$. \textbf{if} $i$ wins, $v_{ij} = 1$ \textbf{else} $v_{ij} = 0$.
\State $w_{ij} = w_{ij} + v_{ij}, r_{ij} = r_{ij} + 1, \hat{p}_{ij} = \frac{w_{ij}}{r_{ij}}$.
\EndWhile
\end{algorithmic}
\textbf{if} $\hat{p}_{ij} \le \frac{1}{2}$ \textbf{Output:} winner $j$ \textbf{else} \textbf{Output:} winner $i$
\end{algorithm}

\subsection{Online learning in crowdsourcing environment}
Online learning can not be readily applied to a crowdsourcing environment for two reasons. First, crowdsourcing requires parallel and asynchronous evaluations, whereas online learning is designed as a sequential algorithm. In crowdsourcing, many workers evaluate the same evaluation target in parallel, and the evaluator's scores are not immediately responded back to an evaluation server after presenting a target pair. Second, experiments using crowdsourcing are designed in fixed budget conditions, whereas online learning does not ensure a fixed number of evaluations for convergence.

We modify MRA to support parallel and asynchronous execution and fixed budget setting for its application to a subjective evaluation in a crowdsourcing environment. Alg.~\ref{alg:nmr} shows the modified MERGE-RANK algorithm. The algorithm is constructed as a state machine whose transitions are triggered by events exchanged with human evaluators. The main modification is about COMPARE: instead of using while-loop-based COMPARE, we replace it with asynchronous execution in any order. The evaluation order of target pairs does not affect the final results because MERGE is associative: $\mathrm{MERGE}((\mathrm{MERGE}(S_1,S_2),S_3) = \mathrm{MERGE}(S_1,\mathrm{MERGE}(S_2,S_3))$. In addition, the divide-and-conquer nature of MRA enables parallel execution easily: every event handler in Alg.~\ref{alg:nmr} can be executed independently in parallel. In concurrent execution, reading and writing states are not atomic, which may cause conflicted state updates. We avoid the issue by using conflict-free incremental operations to update the statistics of the evaluator's scores: $w_{ij} = w_{ij} + v_{ij}, r_{ij} = r_{ij} + 1$. The other statistics, such as $\hat{p}_{ij}, \hat{c}(r_{ij})$ are also conflict-free because they are functions of the conflict-free state values.

Algorithm~\ref{alg:nmr} also supports fixed budget setting. The modified MRA continues evaluation up to budget even after convergence, whereas the MERGE-RANK stops evaluation at convergence. After convergence, the modified MERGE-RANK continues execution to optimize the upper bound of evaluation accuracy of sorted pairs by selecting pairs with the worst error bias to be evaluated. This behavior indicates the algorithm has two phases: before convergence, the algorithm finds optimal pairs and their evaluation volumes by sorting with comparisons that satisfy error bias less than the specified tolerance bias; after convergence, the algorithm further minimizes the error bias for the fixed pairs that are already sorted. In fixed budget conditions, the modified MRA is not guaranteed to converge depending on the value of tolerance bias $\epsilon$. To ensure convergence, the tolerance bias $\epsilon$ has to be configured so that the upper bound of evaluation volume for convergence $m \times \max T(|S|)$ becomes less than the budget.

\begin{figure}[h]
    \begin{center}
    {\includegraphics[width=1.0\columnwidth]{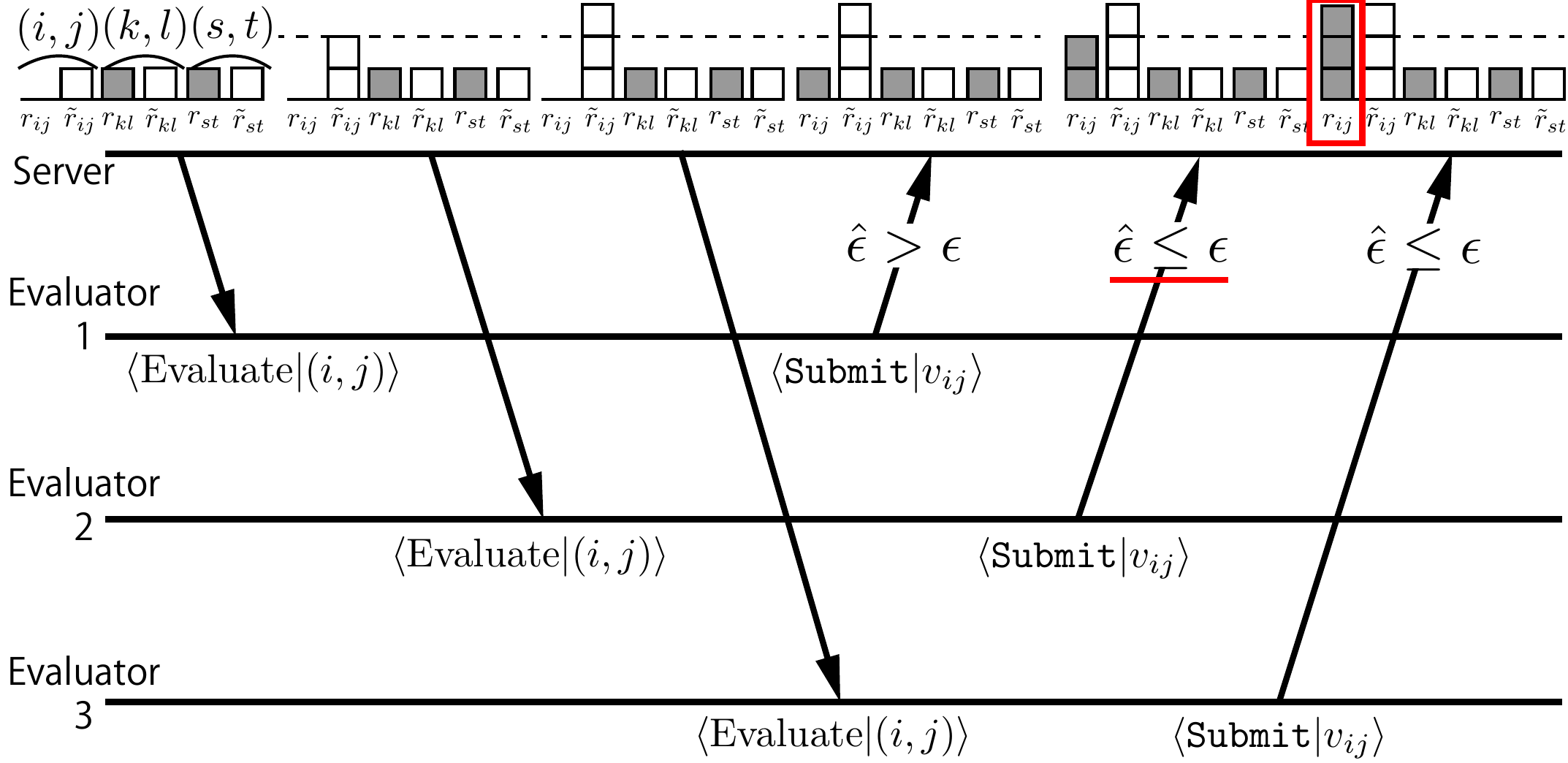}}
    \end{center}
    \caption{Problems for online learning in asynchronous execution. This figure shows a situation where over-evaluation and unbalanced evaluation allocation occurs.}
    \label{fig:asynchronous}
\end{figure}

The asynchronous environment of crowdsourcing causes two challenges in online learning. The first challenge is suboptimal evaluation volume: target pairs may be evaluated more than the optimal volume estimated by the MRA. The second challenge is unbalanced evaluation allocation: a naive implementation selecting a pair with the worst evaluation accuracy as a request to evaluators causes an unbalanced evaluation allocation. These problems are caused by the uncertainty of the evaluation accuracy of target pairs with unreceived scores in asynchronous execution. 
Fig.~\ref{fig:asynchronous} shows an example situation that causes the two challenges. This example shows a situation where three evaluators join simultaneously. The server implementing the MRA returns a \texttt{Evaluate} request selecting a pair from three pairs $(i,j), (k,l), (s,t)$ among which the pair $(i,j)$ has the worst error bias because of zero evaluations. The server selects the same pair $(i,j)$ for all three evaluators because the pair $(i,j)$ remains the worst error bias until the evaluators return their scores. After all the evaluators submit scores, it results in the unbalanced allocation where the pair $(i,j)$ is evaluated three times $r_{ij} = 3$ (marked in the red box where $r_{ij}$ is represented as gray boxes) whereas the other pairs are not evaluated. Fig.~\ref{fig:asynchronous} also shows the suboptimal evaluation volume for a pair $(i,j)$. The MRA decides to stop evaluation for the pair $(i,j)$ at $r_{ij} = 2$ based on the statistical test $\hat{\epsilon} \le \epsilon$ (marked in the red line). Because \texttt{Evaluate} requests are already issued to the three evaluators, it results in over-evaluation with $r_{ij} = 3$ (marked in the red box).

Algorithm~\ref{alg:nmr} implements an additional feature for requesting evaluations to evaluators to mitigate the two asynchronous difficulties. Alg.~\ref{alg:nmr} at line \ref{alg:nmr:tr} manages a requested evaluation count $\tilde{r}_{ij}$ for a pair $(i,j)$ at triggering \texttt{Evaluate} events. The requested evaluation count is used to calculate the expected error bias $\hat{\epsilon}(\tilde{r}_{ij},\hat{p}_{ij})$ by using $\tilde{r}_{ij}$ at line \ref{alg:nmr:ee} in Alg.~\ref{alg:nmr}. Alg.~\ref{alg:nmr} can mitigate the unbalanced evaluation allocation problem by using the expected error bias for pair selection instead of the error bias calculated with evaluated counts $r$. A pair selection reflecting the requested evaluation counts $\tilde{r}$ can distinguish the future error bias at the evaluator's \texttt{Join} event ($\tilde{r}$ displayed as white boxes in Fig.~\ref{fig:asynchronous} is incremented at \texttt{Join} events). The expected error bias reflects a narrowing of the confidence interval based on the incremented requested evaluation counts even though the mean score $\hat{p}_{ij}$ is uncertain due to unreceived evaluations. The balancing mechanism based on the requested evaluation count can also mitigate the suboptimal evaluation volume. The over-evaluation can be avoided by reducing the number of concurrent evaluators joining and evaluating the same pair at the same period. For the example in Fig.~\ref{fig:asynchronous}, Alg.~\ref{alg:nmr} using the balancing mechanism based on the requested evaluation count can distribute \texttt{Evaluate} events for a pair $(i,j), (k,l)$, and $(s,t)$ to evaluator 1, 2, and 3, respectively, instead of distributing only \texttt{Evaluate} events for the same pair $(i,j)$ to them.

\begin{algorithm}[t]
\caption{Modified MERGE-RANK}\label{alg:nmr}
\textbf{Input:} Set $S$, bias $\epsilon$, confidence $\delta$.\\
\textbf{Initialize:} Initialize MERGE-RANK with $S$ which determines target pairs $(i,j)$. $\tilde{r}_{ij} = 0$.\\
\textbf{Events:}\\
- \textbf{Request:} $\langle\mathtt{Join}\rangle$: a join request from an evaluator.\\
- \textbf{Request:} $\langle\mathtt{Submit}|v_{ij}\rangle$: A submission of a preference score $v_{ij} \in \{0,1\}$ for a pair $(i,j)$ from an evaluator.\\
- \textbf{Indication:} $\langle\mathtt{Evaluate}|(i,j)\rangle$: an evaluation request for a pair $(i,j)$ to an evaluator.\\
- \textbf{Indication:} $\langle\mathtt{JoinAgain}\rangle$: a join request to an evaluator.
\begin{algorithmic}[1]
\State\textbf{upon event} $\langle\mathtt{Join}\rangle$ from an evaluator \textbf{do}
\State \hskip2em \textbf{if} there is remaining budget \textbf{do}
\State \hskip4em Select a pair $(i,j)$ with maximum $\hat{\epsilon}(\tilde{r}_{ij},\hat{p}_{ij})$. \label{alg:nmr:ee}
\State \hskip4em $\tilde{r}_{ij} = \tilde{r}_{ij} + 1$. \label{alg:nmr:tr}
\State \hskip4em \textbf{trigger} $\langle\mathtt{Evaluate}|(i,j)\rangle$ to the evaluator.
\State\textbf{upon event} $\langle\mathtt{Submit}|v_{ij}\rangle$ from an evaluator \textbf{do}
\State \hskip2em $w_{ij} = w_{ij} + v_{ij}, r_{ij} = r_{ij} + 1, \hat{p}_{ij} = \frac{w_{ij}}{r_{ij}}$.
\State \hskip2em \textbf{if} $\epsilon \ge \hat{\epsilon}(r_{ij},\hat{p}_{ij})$ or $r_{ij} \ge m$ \textbf{do}
\State \hskip4em \textbf{if} $\hat{p}_{ij} \le \frac{1}{2}$ \textbf{do} $o=j$ \textbf{else} $o=i$.
\State \hskip4em Update MERGE and MERGE-RANK partially 
\State \hskip4em with $o = \mathrm{COMPARE}(i,j,\epsilon,\delta)$ exactly once.
\State \hskip2em \textbf{trigger} $\langle\mathrm{JoinAgain}\rangle$ to the evaluator.
\end{algorithmic}
\end{algorithm}

\section{Experimental evaluation}

\begin{table*}[t]
\caption{Results from the modified MERGE-RANK algorithm. Pairs with statistically significant preference based on the binomial test are marked with ``*". $\underline{r}_{ij}$ is an evaluation volume at the winner determination, and $\bar{r}_{ij}$ is the final evaluation volume. $\underline{p}_{ij}$ is a preference score at the winner determination, and $\bar{p}_{ij}$ is the final preference score. Pairs whose winner is determined by early termination and their related metrics are colored in red. Pairs whose winner is determined by maximum evaluation and their related metrics are colored in blue. Preference scores indicating winner reversal are colored in yellow.}
\label{tbl:scores_svcc2023_jp}
\begin{subtable}[t]{0.45\textwidth}
\begin{center}
\footnotesize
\begin{tabular}{|lrrccccrr|}\hline
$i:j$&$\bar{r}_{ij}$&$\underline{r}_{ij}$&$\bar{p}_{ij}$&$\underline{p}_{ij}$&$\hat{c}$&$\hat{c}_{\mathrm{H}}$&$\hat{\epsilon}$&$\hat{\epsilon}_{\mathrm{H}}$\\\hline
\cellcolor{red!25}TAR:T23*&68&66&0.26&0.27&\cellcolor{red!25}0.31&0.16&\cellcolor{red!25}0.07&-0.07\\
\cellcolor{red!25}SOU:T23*&120&94&0.31&0.32&\cellcolor{red!25}0.24&0.12&\cellcolor{red!25}0.05&-0.07\\
\cellcolor{red!25}T23:T06*&292&199&0.62&0.61&\cellcolor{red!25}0.16&0.08&\cellcolor{red!25}0.05&-0.04\\
\cellcolor{cyan!25}T07:T06*&278&240&0.41&0.42&0.17&\cellcolor{cyan!25}0.08&0.08&\cellcolor{cyan!25}-0.01\\
\cellcolor{red!25}TAR:T06*&184&158&0.35&0.37&\cellcolor{red!25}0.20&0.10&\cellcolor{red!25}0.05&-0.05\\
\cellcolor{red!25}T23:T20*&322&83&0.67&0.70&\cellcolor{red!25}0.16&0.08&\cellcolor{red!25}-0.01&-0.09\\
\cellcolor{cyan!25}T06:T20&503&240&0.51&0.55&0.13&\cellcolor{cyan!25}0.06&0.11&\cellcolor{cyan!25}0.05\\
\cellcolor{cyan!25}T07:T20*&479&240&0.43&0.42&0.13&\cellcolor{cyan!25}0.06&0.06&\cellcolor{cyan!25}-0.01\\
\cellcolor{cyan!25}T02:T20&370&240&0.46&0.47&0.15&\cellcolor{cyan!25}0.07&0.10&\cellcolor{cyan!25}0.03\\
\cellcolor{red!25}T14:T20*&117&103&0.32&0.33&\cellcolor{red!25}0.24&0.13&\cellcolor{red!25}0.07&-0.05\\
\cellcolor{red!25}TAR:T07*&217&180&0.40&0.38&\cellcolor{red!25}0.19&0.09&\cellcolor{red!25}0.08&-0.01\\
\cellcolor{cyan!25}T20:T16*&317&240&0.55&0.55&0.16&\cellcolor{cyan!25}0.08&0.11&\cellcolor{cyan!25}0.02\\
\cellcolor{cyan!25}T07:T16&541&240&\cellcolor{yellow!25}0.49&\cellcolor{yellow!25}0.52&0.13&\cellcolor{cyan!25}0.06&0.11&\cellcolor{cyan!25}0.05\\
\cellcolor{cyan!25}TAR:T16*&440&240&0.44&0.48&0.14&\cellcolor{cyan!25}0.06&0.08&\cellcolor{cyan!25}0.00\\
\cellcolor{cyan!25}T02:T16*&300&240&0.42&0.42&0.16&\cellcolor{cyan!25}0.08&0.09&\cellcolor{cyan!25}0.00\\
\cellcolor{cyan!25}T08:T16*&242&240&0.41&0.41&0.18&\cellcolor{cyan!25}0.09&0.09&\cellcolor{cyan!25}-0.00\\
\cellcolor{red!25}T23:T09*&120&51&0.78&0.76&\cellcolor{red!25}0.24&0.12&\cellcolor{red!25}-0.03&-0.15\\
\cellcolor{red!25}T06:T09*&226&169&0.64&0.62&\cellcolor{red!25}0.18&0.09&\cellcolor{red!25}0.05&-0.05\\
\cellcolor{red!25}T20:T09*&128&80&0.66&0.70&\cellcolor{red!25}0.23&0.12&\cellcolor{red!25}0.08&-0.04\\
\cellcolor{cyan!25}T07:T09*&365&240&0.55&0.51&0.15&\cellcolor{cyan!25}0.07&0.10&\cellcolor{cyan!25}0.03\\
\cellcolor{red!25}T16:T09*&264&238&0.60&0.59&\cellcolor{red!25}0.17&0.08&\cellcolor{red!25}0.07&-0.01\\
\cellcolor{cyan!25}TAR:T09*&274&240&0.57&0.53&0.17&\cellcolor{cyan!25}0.08&0.10&\cellcolor{cyan!25}0.01\\
\cellcolor{cyan!25}T09:T12&464&240&\cellcolor{yellow!25}0.48&\cellcolor{yellow!25}0.50&0.13&\cellcolor{cyan!25}0.06&0.11&\cellcolor{cyan!25}0.04\\
\cellcolor{cyan!25}TAR:T12&620&240&0.51&0.52&0.12&\cellcolor{cyan!25}0.05&0.10&\cellcolor{cyan!25}0.04\\
\cellcolor{cyan!25}T16:B01*&262&240&0.60&0.59&0.17&\cellcolor{cyan!25}0.08&0.08&\cellcolor{cyan!25}-0.01\\
\cellcolor{cyan!25}TAR:B01*&427&240&0.54&0.57&0.14&\cellcolor{cyan!25}0.07&0.10&\cellcolor{cyan!25}0.02\\
\cellcolor{red!25}T02:B01&331&18&\cellcolor{yellow!25}0.54&\cellcolor{yellow!25}0.06&\cellcolor{red!25}0.16&0.07&\cellcolor{red!25}0.11&0.03\\
\cellcolor{cyan!25}T13:B01&397&240&0.47&0.48&0.14&\cellcolor{cyan!25}0.07&0.11&\cellcolor{cyan!25}0.04\\
\cellcolor{cyan!25}T21:B01&541&240&\cellcolor{yellow!25}0.51&\cellcolor{yellow!25}0.46&0.13&\cellcolor{cyan!25}0.06&0.11&\cellcolor{cyan!25}0.05\\
\cellcolor{red!25}T08:B01*&246&213&0.39&0.40&\cellcolor{red!25}0.18&0.09&\cellcolor{red!25}0.06&-0.03\\
\cellcolor{red!25}T13:T02*&361&232&0.43&0.41&\cellcolor{red!25}0.15&0.07&\cellcolor{red!25}0.08&-0.00\\
\cellcolor{red!25}T06:T13*&102&99&0.68&0.68&\cellcolor{red!25}0.26&0.13&\cellcolor{red!25}0.08&-0.04\\
\cellcolor{red!25}T07:T13*&193&158&0.63&0.63&\cellcolor{red!25}0.20&0.10&\cellcolor{red!25}0.06&-0.03\\
\cellcolor{cyan!25}TAR:T13&470&240&0.54&0.55&0.13&\cellcolor{cyan!25}0.06&0.10&\cellcolor{cyan!25}0.03\\
\cellcolor{cyan!25}SOU:T13*&484&240&0.46&0.45&0.13&\cellcolor{cyan!25}0.06&0.09&\cellcolor{cyan!25}0.02\\
\cellcolor{cyan!25}T12:T01*&269&240&0.59&0.59&0.17&\cellcolor{cyan!25}0.08&0.08&\cellcolor{cyan!25}-0.01\\
\cellcolor{red!25}TAR:T01*&226&161&0.64&0.63&\cellcolor{red!25}0.18&0.09&\cellcolor{red!25}0.05&-0.05\\
\cellcolor{red!25}B01:T01*&179&137&0.63&0.64&\cellcolor{red!25}0.20&0.10&\cellcolor{red!25}0.08&-0.02\\
\cellcolor{red!25}T02:T01*&114&61&0.70&0.74&\cellcolor{red!25}0.25&0.13&\cellcolor{red!25}0.04&-0.07\\
\cellcolor{cyan!25}T13:T01*&278&240&0.56&0.57&0.17&\cellcolor{cyan!25}0.08&0.10&\cellcolor{cyan!25}0.02\\
\cellcolor{cyan!25}T21:T01*&277&240&0.42&0.42&0.17&\cellcolor{cyan!25}0.08&0.08&\cellcolor{cyan!25}-0.00\\
\cellcolor{cyan!25}T02:T21*&289&240&0.57&0.59&0.16&\cellcolor{cyan!25}0.08&0.09&\cellcolor{cyan!25}0.01\\
\hline
\end{tabular}
\end{center}
\end{subtable}
\hspace{8mm}
\begin{subtable}[t]{0.45\textwidth}
\begin{center}
\footnotesize
\begin{tabular}{|lrrccccrr|}\hline
$i:j$&$\bar{r}_{ij}$&$\underline{r}_{ij}$&$\bar{p}_{ij}$&$\underline{p}_{ij}$&$\hat{c}$&$\hat{c}_{\mathrm{H}}$&$\hat{\epsilon}$&$\hat{\epsilon}_{\mathrm{H}}$\\\hline
\cellcolor{cyan!25}T13:T21&590&240&\cellcolor{yellow!25}0.49&\cellcolor{yellow!25}0.55&0.12&\cellcolor{cyan!25}0.06&0.11&\cellcolor{cyan!25}0.05\\
\cellcolor{cyan!25}SOU:T21&348&240&0.46&0.45&0.15&\cellcolor{cyan!25}0.07&0.11&\cellcolor{cyan!25}0.04\\
\cellcolor{cyan!25}T08:T21&527&240&0.49&0.50&0.13&\cellcolor{cyan!25}0.06&0.11&\cellcolor{cyan!25}0.04\\
\cellcolor{cyan!25}T14:T21&426&240&0.47&0.48&0.14&\cellcolor{cyan!25}0.07&0.11&\cellcolor{cyan!25}0.03\\
\cellcolor{cyan!25}TAR:SOU&483&240&\cellcolor{yellow!25}0.48&\cellcolor{yellow!25}0.53&0.13&\cellcolor{cyan!25}0.06&0.11&\cellcolor{cyan!25}0.04\\
\cellcolor{cyan!25}SOU:T08*&313&240&0.55&0.55&0.16&\cellcolor{cyan!25}0.08&0.11&\cellcolor{cyan!25}0.02\\
\cellcolor{cyan!25}T14:T08*&369&240&0.44&0.44&0.15&\cellcolor{cyan!25}0.07&0.09&\cellcolor{cyan!25}0.02\\
\cellcolor{red!25}T02:T14*&135&119&0.65&0.66&\cellcolor{red!25}0.23&0.12&\cellcolor{red!25}0.08&-0.03\\
\cellcolor{red!25}T09:T22*&189&167&0.62&0.62&\cellcolor{red!25}0.20&0.10&\cellcolor{red!25}0.07&-0.03\\
\cellcolor{red!25}T12:T22*&160&128&0.65&0.65&\cellcolor{red!25}0.21&0.11&\cellcolor{red!25}0.06&-0.04\\
\cellcolor{cyan!25}T01:T22*&308&240&0.56&0.56&0.16&\cellcolor{cyan!25}0.08&0.10&\cellcolor{cyan!25}0.02\\
\cellcolor{cyan!25}T21:T22*&280&240&0.60&0.59&0.17&\cellcolor{cyan!25}0.08&0.07&\cellcolor{cyan!25}-0.02\\
\cellcolor{cyan!25}SOU:T22*&278&240&0.60&0.59&0.17&\cellcolor{cyan!25}0.08&0.07&\cellcolor{cyan!25}-0.02\\
\cellcolor{cyan!25}T08:T22&623&240&\cellcolor{yellow!25}0.49&\cellcolor{yellow!25}0.53&0.12&\cellcolor{cyan!25}0.05&0.11&\cellcolor{cyan!25}0.05\\
\cellcolor{cyan!25}T14:T22*&293&240&0.55&0.56&0.16&\cellcolor{cyan!25}0.08&0.11&\cellcolor{cyan!25}0.03\\
\cellcolor{cyan!25}T11:T22*&253&240&0.44&0.44&0.17&\cellcolor{cyan!25}0.09&0.11&\cellcolor{cyan!25}0.02\\
\cellcolor{cyan!25}T17:T11&303&240&0.45&0.45&0.16&\cellcolor{cyan!25}0.08&0.11&\cellcolor{cyan!25}0.03\\
\cellcolor{red!25}T09:T19*&151&28&0.89&0.86&\cellcolor{red!25}0.22&0.11&\cellcolor{red!25}-0.17&-0.28\\
\cellcolor{red!25}T12:T19*&152&14&0.88&1.00&\cellcolor{red!25}0.22&0.11&\cellcolor{red!25}-0.16&-0.27\\
\cellcolor{red!25}T01:T19*&59&25&0.86&0.88&\cellcolor{red!25}0.33&0.18&\cellcolor{red!25}-0.04&-0.19\\
\cellcolor{red!25}T22:T19*&62&22&0.90&0.91&\cellcolor{red!25}0.32&0.17&\cellcolor{red!25}-0.08&-0.23\\
\cellcolor{red!25}T11:T19*&137&25&0.79&0.88&\cellcolor{red!25}0.23&0.12&\cellcolor{red!25}-0.06&-0.17\\
\cellcolor{red!25}T17:T19*&87&18&0.90&0.94&\cellcolor{red!25}0.28&0.15&\cellcolor{red!25}-0.12&-0.25\\
\cellcolor{cyan!25}T15:T19&466&240&0.48&0.49&0.13&\cellcolor{cyan!25}0.06&0.11&\cellcolor{cyan!25}0.04\\
\cellcolor{red!25}T24:T19*&232&188&0.35&0.39&\cellcolor{red!25}0.18&0.09&\cellcolor{red!25}0.03&-0.06\\
\cellcolor{red!25}T22:T15*&30&28&0.87&0.86&\cellcolor{red!25}0.43&0.25&\cellcolor{red!25}0.07&-0.12\\
\cellcolor{red!25}T11:T15*&70&68&0.73&0.72&\cellcolor{red!25}0.30&0.16&\cellcolor{red!25}0.07&-0.07\\
\cellcolor{red!25}T17:T15*&139&37&0.81&0.81&\cellcolor{red!25}0.23&0.12&\cellcolor{red!25}-0.08&-0.19\\
\cellcolor{cyan!25}T19:T05&318&240&0.54&0.53&0.16&\cellcolor{cyan!25}0.08&0.11&\cellcolor{cyan!25}0.03\\
\cellcolor{red!25}T15:T05*&210&209&0.60&0.60&\cellcolor{red!25}0.19&0.09&\cellcolor{red!25}0.08&-0.01\\
\cellcolor{cyan!25}T18:T05&563&240&0.49&0.48&0.12&\cellcolor{cyan!25}0.06&0.11&\cellcolor{cyan!25}0.04\\
\cellcolor{cyan!25}T10:T05&411&240&0.47&0.47&0.14&\cellcolor{cyan!25}0.07&0.11&\cellcolor{cyan!25}0.04\\
\cellcolor{cyan!25}T19:T18&663&240&0.50&0.50&0.11&\cellcolor{cyan!25}0.05&0.11&\cellcolor{cyan!25}0.05\\
\cellcolor{cyan!25}T24:T18*&372&240&0.45&0.44&0.15&\cellcolor{cyan!25}0.07&0.10&\cellcolor{cyan!25}0.02\\
\cellcolor{cyan!25}T18:T10*&450&240&0.56&0.53&0.14&\cellcolor{cyan!25}0.06&0.07&\cellcolor{cyan!25}0.00\\
\cellcolor{red!25}T24:T10*&327&156&0.38&0.37&\cellcolor{red!25}0.16&0.08&\cellcolor{red!25}0.04&-0.04\\
\cellcolor{cyan!25}T03:T10*&310&240&0.45&0.47&0.16&\cellcolor{cyan!25}0.08&0.11&\cellcolor{cyan!25}0.03\\
\cellcolor{red!25}T05:B02*&114&108&0.66&0.67&\cellcolor{red!25}0.25&0.13&\cellcolor{red!25}0.09&-0.03\\
\cellcolor{red!25}T10:B02*&229&128&0.63&0.65&\cellcolor{red!25}0.18&0.09&\cellcolor{red!25}0.05&-0.04\\
\cellcolor{cyan!25}T24:B02*&359&240&0.56&0.56&0.15&\cellcolor{cyan!25}0.07&0.09&\cellcolor{cyan!25}0.01\\
\cellcolor{cyan!25}T03:B02&474&240&0.53&0.50&0.13&\cellcolor{cyan!25}0.06&0.10&\cellcolor{cyan!25}0.03\\
~&~&~&~&~&~&~&~&~\\
\hline
\end{tabular}
\end{center}
\end{subtable}
\end{table*}

\subsection{Experimental conditions}
We implemented the modified MRA in a listening test server. The listening test server supported various subjective evaluations such as the MOS test, and it could accept evaluations submitted by listeners collected with crowdsourcing. The server could handle large-scale evaluations using crowdsourcing such as international competitions: the Voice Conversion Challenge 2020 \cite{DBLP:journals/corr/abs-2008-12527}, the Voice MOS Challenge 2022 \cite{huang22f_interspeech}, and the Singing Voice Conversion Challenge (SVCC) 2023 \cite{DBLP:journals/corr/abs-2306-14422} were conducted on this server.

We selected 27 speech systems from a main task of the SVCC 2023 \cite{DBLP:journals/corr/abs-2306-14422} as evaluation targets of the dynamic preference test to evaluate the optimization performance of the modified MRA. The total combinations from the 27 systems were 351 pairs. The 27 speech systems were synthetic singing voice systems and natural samples from source and target speakers that were already evaluated in the MOS test about naturalness. Thus, we knew the overall order of their naturalness in terms of MOS. We followed the same listening test content as SVCC2023: we included 96 samples, which consisted of 24 utterances from two source and two target speaker combinations for each system as evaluation targets.

We designed the preference test as 416 sets with 60 pages. The total budget for the test was thus 24,960 evaluations. Each page contained two samples, and a listener was asked to select a sample that sounded more natural than the other. The contents of the test, which are about pair combinations and evaluation volumes for a pair, were automatically designed by the modified MRA online based on the listener's evaluations. The speech samples presented to listeners were selected from a queue for a pair, which ensures a balanced sample selection. We collected 321 Japanese listeners by crowdsourcing.

We configured the modified MRA with an initial order of the 27 systems sorted based on the MOS. The sorting complexity of MRA depends on the initial order of input $S$, whose number of compared pairs to converge ranges between $60 \le T(|S|) \le 104$ when $|S| = 27$ \footnote{Not that the sorting complexity should be computed recursively when $|S| \ne 2^k$, by using the lower bound $T(n) \ge T(\lceil n/2 \rceil) + T(\lfloor n/2 \rfloor) + \lfloor \frac{n}{2}\rfloor$ and the upper bound $T(|S|) \le T(\lceil n/2 \rceil) + T(\lfloor n/2 \rfloor) + n - 1$.}. We used the MOS-based initial order for two reasons: (1) to avoid the worst sorting complexity and (2) to investigate how much the sorting order based on preference differed from the sorting order based on MOS. We configured the tolerance bias as $\epsilon = 0.0877$ and confidence as $\delta = 0.05$. This configuration meant the maximum evaluation volume to determine a winner from each pair was $m=240$. In these configurations, the worst evaluation volume was bounded between 14,400 and 24,960, where all pairs were evaluated up to the maximum counts $m=240$. The worst complexity of 24,960 was equal to our budget of 24,960, which ensured the convergence of the MRA in the experiment.

We also compared the preference scores with pre-recorded MOS in the SVCC2023. To do so, we analyzed results obtained from the modified MRA with standard statistical analysis methods. We tested the statistical significance of the final preference scores with the one-sided binomial test if preferences were less or greater than 0.5 with 95\% confidence. We calculated the Clopper–Pearson confidence interval for the preference scores with 95\% confidence \cite{be7c0fd0-f562-39ad-b8e0-716a276561d1}. We used MOS from both English and Japanese evaluators in the same task, which contained 13,440 scores in total that broke down to 3,360 evaluations from English listeners and 10,080 evaluations from Japanese listeners. For a statistical test of MOS, we used the one-sided Mann-Whitney U-test with 95\% confidence \cite{Mann-WhitneyRankTest} and 95\% confidence interval based on the central limit theorem with Student-t distribution. Note that the purpose of the binomial test and the Clopper–Pearson confidence interval was different from the statistical test and confidence intervals used for the MRA. The purpose of the statistical test and confidence intervals used in the MRA is to ensure evaluation accuracy under dynamically optimized evaluation allocation in uncertainty. Thus, the modified MRA uses conservative statistical tests and confidence intervals based on Hoeffding's inequality that provides an upper bound. After the evaluation, our interest is in precise statistical tests and confidence intervals for fixed results. 

\subsection{Experimental results}
\begin{figure}[t]
    \begin{center}
    {\includegraphics[width=1.0\columnwidth]{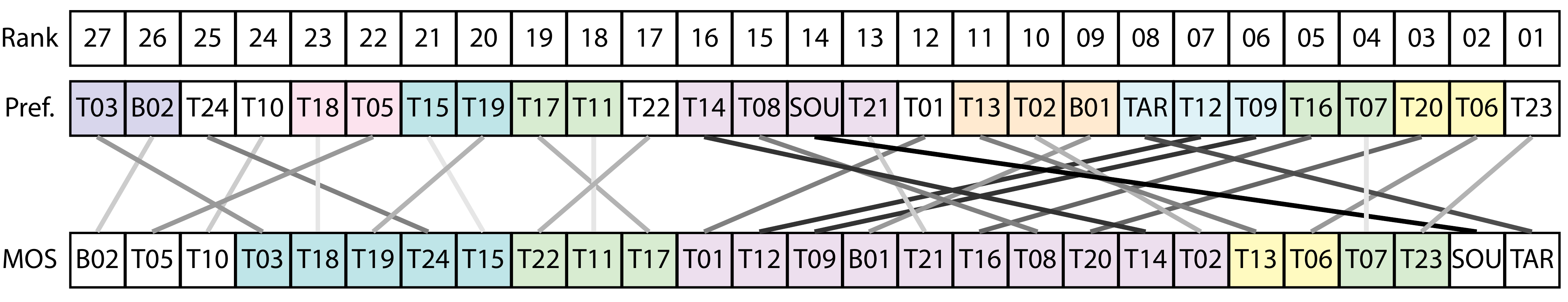}}
    \end{center}
    \caption{The final and initial order obtained with the modified MRA in a preference test. The initial order is based on the pre-recorded MOS. Adjacent pairs without statistical significance are colored in the same color.}
    \label{fig:order-comparison}
\end{figure}

Table~\ref{tbl:scores_svcc2023_jp} shows the results of the dynamic preference test. The number of evaluated pairs was 83, which was much lower than the total combination of 351. The number of evaluated pairs was slightly higher than the minimum sorting complexity 60. This indicated that the final order based on preference slightly differed from the initial order based on MOS. 
The evaluation allocations were divided into two phases: the sorting phase before convergence and the remaining budget consumption phase after convergence. The MRA converged at 15,248 evaluations to sort all 27 systems. 9,712 evaluations were the remaining budget, and the MRA consumed them to reduce the worst error bias. 
Table~\ref{tbl:scores_svcc2023_jp} reflects the two phases of the evaluation in evaluation volumes and preference scores: $\underline{r}_{ij}$ is an evaluation volume at the winner determination from the first phase, and $\bar{r}_{ij}$ is the final evaluation volume from the second phase; $\underline{p}_{ij}$ is a preference score at the winner determination from the first phase and $\bar{p}_{ij}$ is the final preference score from the second phase. For example, The pair T12:T19 was allocated only $\underline{r}_{ij}=14$ evaluations to determine T12 as a winner. This small allocation was because the pair T12:T19 had distinct preference $\underline{p}_{ij} = 1.0$. This pair was evaluated up to $\bar{r}_{ij}=152$ times after convergence. For another example, the pair T19:T18 was evaluated up to the maximum limit $\underline{r}_{ij}=m=240$ to determine T19 as a winner. The pair T19:T18 was further evaluated up to $\bar{r}_{ij}=663$ times. This large allocation was because this pair had tie preference $\bar{p}_{ij} = 0.5$.

Even though evaluations were not evenly allocated, the evaluation accuracies were not affected. All pairs achieved error bias $\hat{\epsilon}_H$ at most the target tolerance bias $\epsilon=0.0877$. The MRA decided winners by terminating evaluations early based on error bias $\hat{\epsilon}$. There were 36 pairs evaluated in this way, and they were colored in red in Table~\ref{tbl:scores_svcc2023_jp}. Except for one pair, these pairs achieved error bias $\hat{\epsilon}$ at most the target tolerance bias $\epsilon=0.0877$. Only the pair T02:B01 did not achieve error bias less than the tolerance bias, which error bias was $\hat{\epsilon}=0.11$. We will analyze this pair in later sections. The MRA evaluated up to the maximum limit of evaluations $m=240$ to decide winners if the early termination could not be performed. There were 47 pairs evaluated in this way, and they were colored in blue in Table~\ref{tbl:scores_svcc2023_jp}.

\begin{figure}[t!]
    \begin{subfigure}[b]{\columnwidth}
        \centering
        \includegraphics[width=0.86\textwidth]{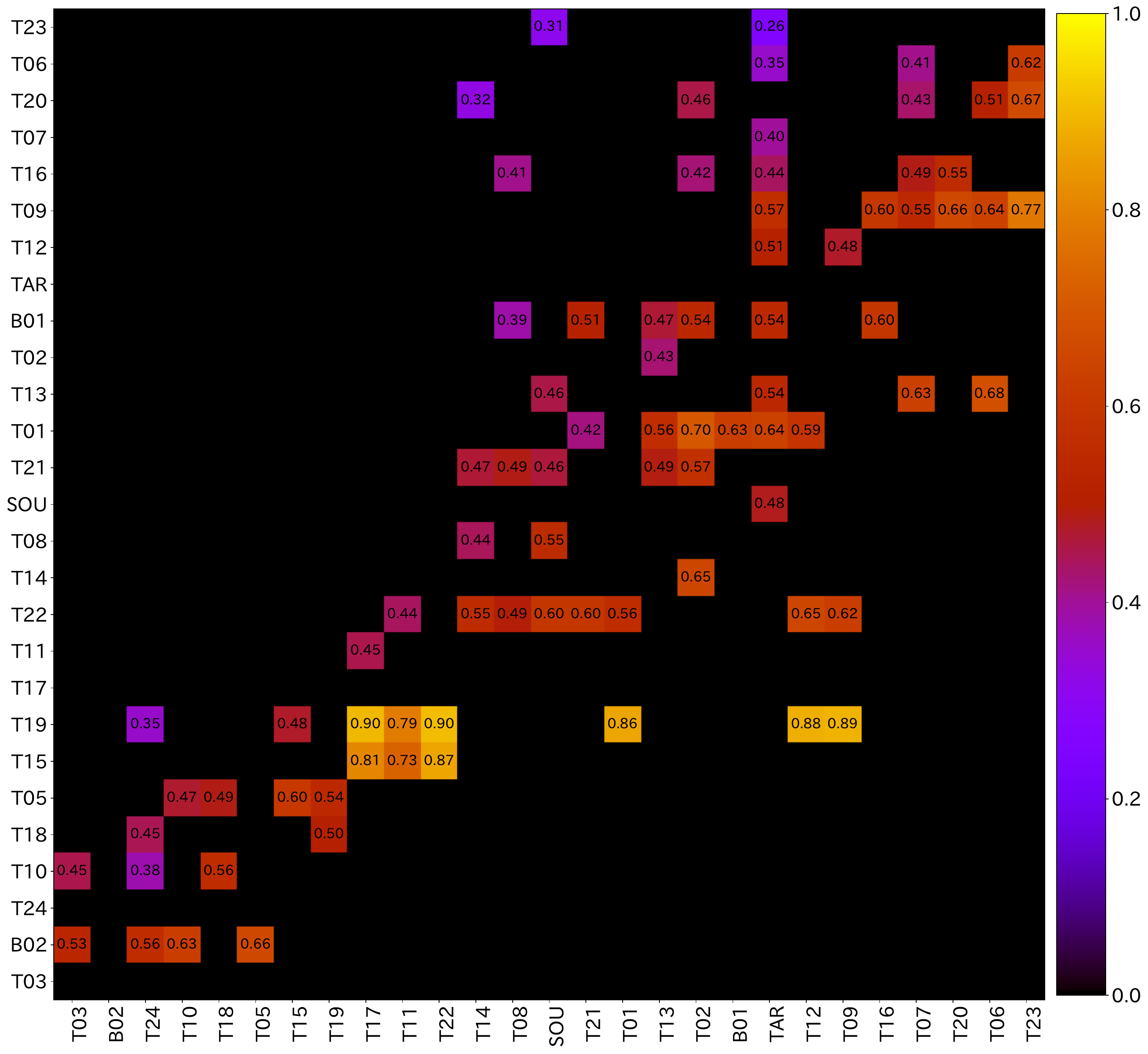}
        \caption{Preference scores $\bar{p}_{ij}$}
        \label{fig:score_distribution:preference}
    \end{subfigure}
    \begin{subfigure}[b]{\columnwidth}
        \centering
        \includegraphics[width=0.86\textwidth]{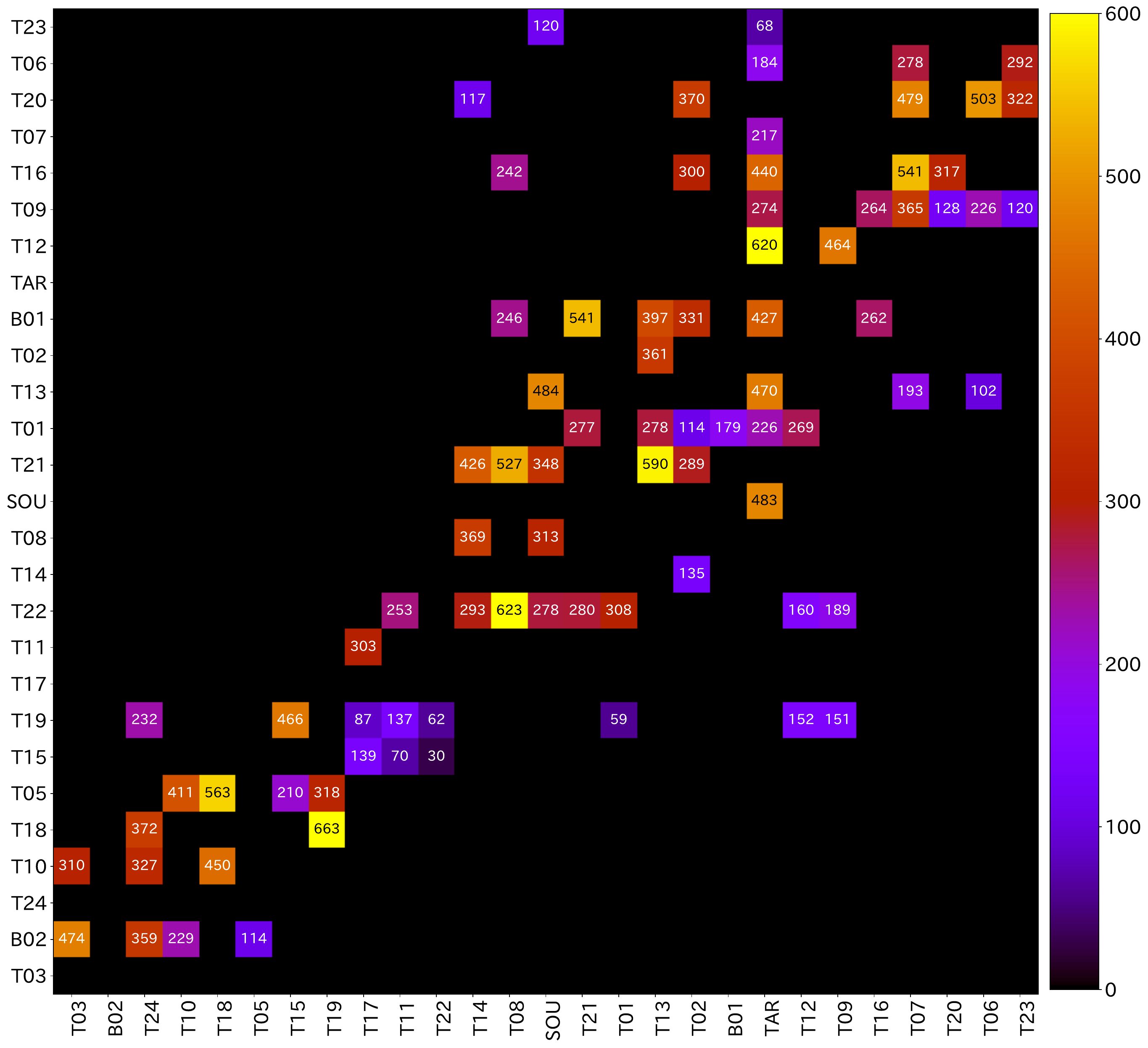}
        \caption{The number of evaluations $\bar{r}_{ij}$}
        \label{fig:score_distribution:n_samples}
    \end{subfigure}
    \begin{subfigure}[b]{\columnwidth}
        \centering
        \includegraphics[width=0.86\textwidth]{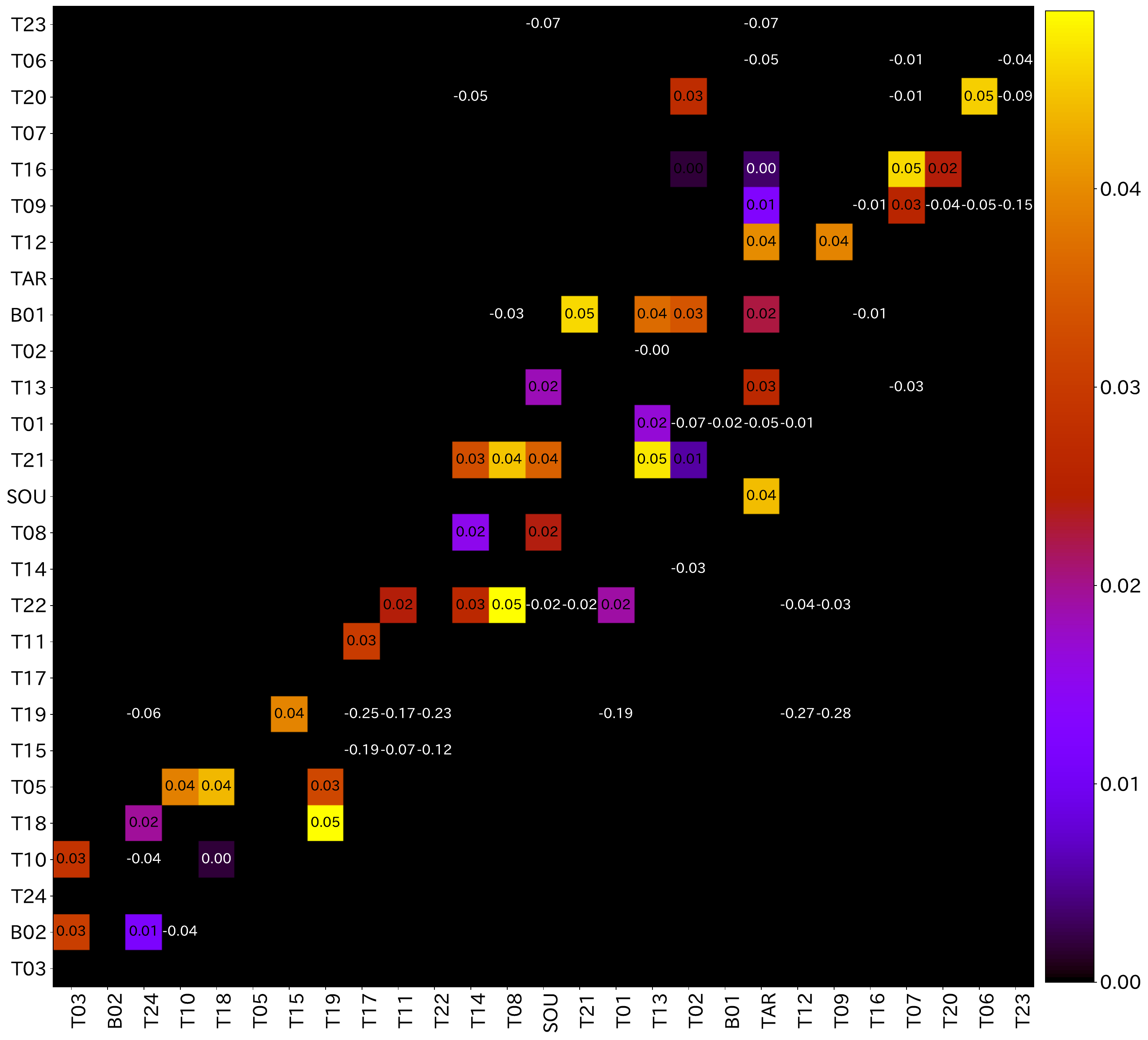}
        \caption{Error bias $\hat{\epsilon}_H$}
        \label{fig:score_distribution:tolerance}
    \end{subfigure}
    \caption{Distributions of metrics from pairwise evaluations.}
    \label{fig:score_distribution}
\end{figure}

Figure~\ref{fig:order-comparison} compares the final order obtained by the modified MRA with the initial order based on the pre-recorded MOS. Adjacent pairs without statistical significance are colored in the same color. In the initial order based on MOS, there was a large region of pairs without statistical significance in the middle ranking. This was caused by the contraction bias of MOS. In contrast, the preference-based test could distinguish quality differences across the whole region, which indicated that the preference test could avoid the contraction bias. Many reordering were observed in the middle ranking. It was because the MOS evaluation did not distinguish quality differences in that region.

Figure~\ref{fig:score_distribution:preference} shows the distribution of preference scores $\bar{p}_{ij}$. Scores of similar preference close to 0.5 were distributed near the diagonal region. Scores of distinct preferences far from 0.5 were distributed apart from the diagonal region. Most pairs in off-diagonal regions were not evaluated. They indicated that the MRA focused on evaluating pairs with similar preferences. 

Figure \ref{fig:score_distribution:n_samples} shows the distribution of the number of evaluations $\bar{r}_{ij}$ for each pair. The distribution of the evaluation volume followed a similar trend to the preference score: large evaluation volumes were distributed in pairs in the diagonal region. This indicated that the MRA prioritized evaluations to pairs with similar preferences. The minimum number of evaluations was 30 of the pair T22:T15. This small allocation of evaluation volume was for the early termination because the pair T22:T15 had a distinct preference of $\bar{p}_{ij}=0.87$. The maximum number of evaluations was 663 of the pair T19:T18. This large allocation of evaluation volume was because the pair T19:T18 had a similar preference of $\bar{p}_{ij}=0.50$. The evaluation volume of 663 from the pair T19:T18 was above the max limit of evaluation to determine a winner $m=240$. This indicated that after convergence, the MRA allocated more evaluations to this pair by consuming the remaining budget.

Figure \ref{fig:score_distribution:tolerance} shows the distribution of error bias. All pairs had error bias at most $\hat{\epsilon}_H = 0.05$, which was smaller than the tolerance bias $\epsilon = 0.0877$. This result showed the property of the algorithm empirically that guaranteed error bias was lower than tolerance bias if it converged. The achieved error bias was much smaller than the tolerance bias, indicating that the algorithm effectively utilized the remaining budget to reduce the error bias further. The figure also shows that pairs with large error bias were distributed in diagonal regions, whereas the pairs with small or no error bias were distributed in off-diagonal regions. This distribution reflected the quality differences of the pairs: there remained large error biases in pairs with similar quality.

\begin{figure}[t]
    \begin{center}
    {\includegraphics[width=1.0\columnwidth]{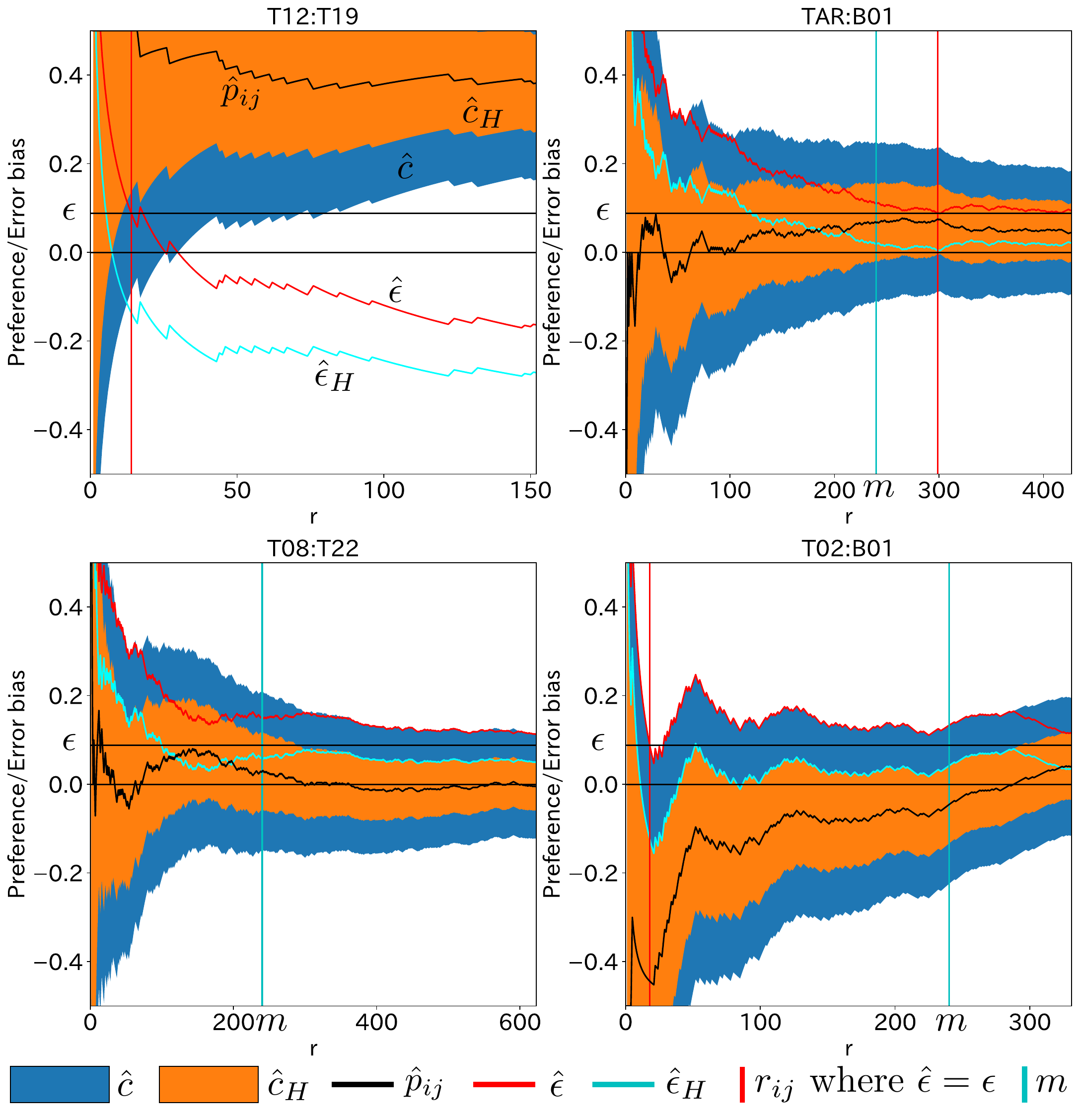}}
    \end{center}
    \caption{Evaluation histories from each pair. The x-axis shows the number of evaluations $r_{ij}$. Centered preference $\hat{p}_{ij} - 0.5$ is represented in the black line. The confidence interval $\hat{c}$ is shown as the blue area, and the confidence interval $\hat{c}_H$ is shown as the orange area. The red line shows error bias $\hat{\epsilon}$, and the cyan line shows error bias $\hat{\epsilon}_H$. The vertical red line shows $r_{ij}$ where the early termination criteria $\hat{\epsilon} \le \epsilon$ is met. Maximum evaluation limit $m$ is shown as a vertical cyan line.}
    \label{fig:score_history_selected}
\end{figure}

Figure \ref{fig:score_history_selected} shows evaluation histories for pairs that experienced representative situations. The pair T12:T19 was an example pair of the early termination. The pair T12:T19 reached $\hat{\epsilon} = \epsilon$ at $\underline{r}_{ij}=14$ before it reached the maximum limit $\underline{r}_{ij} = m$. The MRA determined T12 as a winner based on $\underline{p}_{ij} = 1.0 > 0.5$. After convergence, the error bias of T12:T19 was further reduced up to $\hat{\epsilon} = 0.11$ by allocating $\bar{r}_{ij}=152$. At this end, the preference became $\bar{p}_{ij}=0.88$, which kept T12 as the winner. The pair TAR:B01 was an example of a winner determination by the max limit. The pair TAR:B01 reached the maximum limit $\underline{r}_{ij} = m$ before it reached the early termination criteria $\hat{\epsilon} = \epsilon$. At this point, The pair TAR:B01 had a preference of $\underline{p}_{ij} = 0.57$, and TAR was determined as a winner.
This pair continued to be evaluated up to $\bar{r}_{ij}=427$ after convergence, which resulted in smaller error bias $\hat{\epsilon}_H=0.02$. It still had preference $\bar{p}_{ij}=0.54$ after continuing its evaluation, so the final winner of this pair was unchanged after the convergence. The pair T08:T22 was an example of winner reversal. At the maximum evaluation limit $\underline{r}_{ij}=m$, the MRA determined T08 as a winner based on its preference $\underline{p}_{ij}=0.53$. After continuing evaluation after the convergence, the final winner became opposite to the winner determined at the convergence because the final preference became less than 0.5 as $\bar{p}_{ij}=0.49$. The final preference of $\bar{p}_{ij}=0.49$ could be considered close to a tie, and the preference was not statistically significant, so the negative impact of this winner reversal to the sort order could be negligible in this case. The truly unwanted winner reversal situations could be observed in the pair T02:B01. The MRA terminated the evaluation early at $\underline{r}_{ij}=18$ because it reached the early termination criteria $\hat{\epsilon} = \epsilon$. At this point, the MRA determined B01 as a winner based on its preference $\underline{p}_{ij}=0.06$. However, after continuing its evaluation to $\bar{r}_{ij} = 331$, the winner was changed to the T02 based on its preference of $\bar{p}_{ij}=0.54$. After the convergence, error bias $\hat{\epsilon}$ became larger than the tolerance bias $\epsilon$ as the evaluation continued, and it could not achieve the tolerance bias at last. The first preference $\underline{p}_{ij}=0.06$ and the last preference $\bar{p}_{ij}=0.54$ of this pair were quite different, indicating that the winner estimation by the early termination was unsuccessful. This situation happened with a small probability $\delta = 0.05$ and could be avoided using a smaller error probability. Using larger tolerance bias $\epsilon$ also could prevent the winner reversal. Using safer error probability and tolerance bias means it requires more budget. Thus, we consider only one winner reversal reasonable under our budget limitation.

\begin{figure}[t]
    \begin{center}
    {\includegraphics[width=1.0\columnwidth]{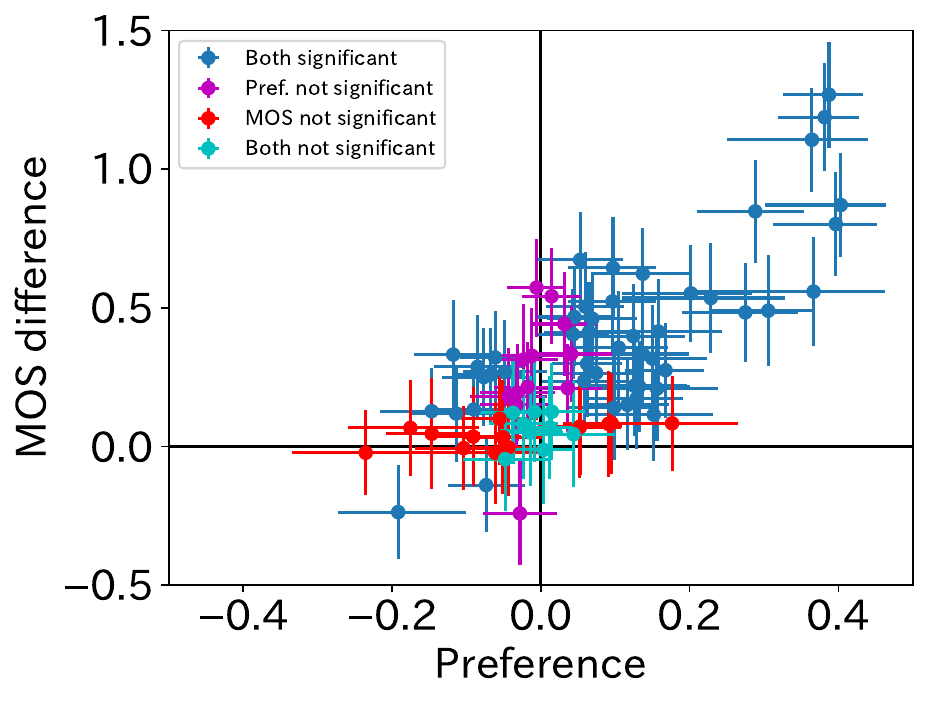}}
    \end{center}
    \caption{Relationship between preference scores and MOS differences. The x-axis shows centered preference $\hat{p}_{ij} - 0.5$.}
    \label{fig:pref_mos_compare_jp}
\end{figure}

Figure~\ref{fig:pref_mos_compare_jp} shows a relationship between preferences and MOS differences. There were 47 pairs where both their preferences and MOS differences were statistically significant. There were 12 pairs whose preferences were not statistically significant, but their MOS differences were statistically significant.
There were 14 pairs whose preferences were statistically significant, but their MOS differences were not statistically significant. There were 10 pairs whose preferences and MOS differences were not statistically significant on both sides. In terms of the number of statistically significant pairs, our dynamic preference test provided better evaluations than MOS: 61 pairs were statistically significant in the preference test, and 59 pairs were statistically significant in the MOS test. Note that the budget for the MOS experiment, which was 13,440 evaluations, was smaller than our preference-based test, which was 24,960 evaluations. To compare the two tests under the same budget level, we measured the number of statistically significant pairs at the convergence of the modified MRA, which was at 15,248 evaluations. 36 pairs were statistically significant in the preference test at the convergence. This indicated that the preference-based test optimized by the modified MRA was not as cost-effective as the MOS evaluation, and further optimization was required to reach the same cost as MOS evaluations.

There was a rough correlation between the preferences and MOS differences. 38 pairs had the same winner relationship between the two evaluation methods, among the 47 pairs where both their preferences and MOS differences were statistically significant, as shown in the first and third quadrants in the figure. Nine pairs had the opposite winner relationship between the two evaluation methods, among the 47 pairs with statistical significance on both sides, as shown in the second and fourth quadrants in the figure. Preference scores are considered more reliable than MOS, so our large-scale preference test could validate MOS tests. Further investigation would be required to verify the MOS by preference tests, in addition to validation of MOS by MOS \cite{cooper2023investigating, LEMAGUER2024101577}.


\section{Conclusion}
\label{sec:conclusion}
This paper optimized preference-based subjective evaluation in terms of pair combination selections and allocation of evaluation volumes with online learning in a crowdsourcing environment. Our algorithm was based on MERGE-RANK, which was a preference-based online learning method based on a sorting algorithm to identify the total order of evaluation targets with minimum sample volumes to achieve specified evaluation accuracy. We modified the MERGE-RANK algorithm to support its execution in a crowdsourcing environment that required parallel and asynchronous execution under fixed budget settings. We introduced a balancing evaluation allocation mechanism to our algorithm to cope with unbalanced and suboptimal evaluation allocation challenges caused by the asynchronous execution of online learning in crowdsourcing. In experiments about the preference-based subjective evaluation of synthetic speech using crowdsourcing, our modified MERGE-RANK algorithm successfully reduced pair combinations from 351 to 83 without compromising evaluation accuracies and wasting budget allocations. The evaluation volume allocation was also optimized: the minimum allocation was 30 evaluations for a pair with distinct quality, and the maximum allocation was 663 evaluations for a pair with similar quality. Only one pair showed a winner reversal after early termination of evaluation, which indicated the obtained total order had an error caused by the stochastic nature of preference tests. In our future work, we will make our algorithm robust to the winner reversal situation more and optimize the efficiency of the algorithm by reducing pairs to be evaluated further.

\section{Acknowledgements}
This work was partly supported by JST CREST Grant Number JPMJCR19A3 including the AIP challenge program.

\bibliographystyle{IEEEbib}
\bibliography{mybib}

\end{document}